# Radiation constraints from cosmic strings

R.A.Battye and E.P.S.Shellard[OT1] [*]

[a]Department of Applied Mathematics and Theoretical Physics, University of Cambridge,
Silver Street, Cambridge, CB3 9EW, U.K.

We show that it is possible to evolve a network of global strings numerically including the effects of radiative backreaction, using the renormalised equations for the Kalb-Ramond action. We calculate radiative corrections to the equations of motion and deduce the effect on a network of global strings. We also discuss the implications of this work for the cosmological axion density.

## 1. INTRODUCTION

Cosmic strings can play many interesting roles in the early universe. It has been shown that a network of cosmic strings will evolve towards a self-similar scaling solution on the largest scales. However, on the smallest scales, where radiative effects are important, the evolution is less well understood. Here we report on an ongoing research project[1-3] to understand the true nature of these effects. Our primary motivation is to calculate accurately the amplitude and spectrum of the radiation background from both local and global strings. For the purposes of this paper we shall concentrate on Goldstone boson radiation from global strings, however analogous calculations have been made for gravitational radiation from local strings[4].

## 2. RADIATIVE BACKREACTION

The starting point for our discussion of global strings will be the Kalb-Ramond action[7]

$$S = -\mu_0 \int d\sigma d\tau \sqrt{-\gamma} - \frac{1}{6} \int d^4x H^2 - 2\pi f_a \int B_{\mu\nu} d\sigma^{\mu\nu}, \quad (1)$$

which has been shown to exhibit the same dynamics as the U(1) Goldstone model[1] using comparisons between numerical field theory simulations and analytic calculations of the radiation power. The antisymmetric tensor field $B_{\mu\nu}$ is dual to the massless Goldstone boson associated with the string, $H_{\mu\alpha\beta}$ is the field strength associated with $B_{\mu\nu}$ and $d\sigma^{\mu\nu} = \epsilon^{ab}\partial_a X^\mu \partial_b X^\nu$ is the area element in terms of the worldsheet coordinates $X^\mu(\sigma,\tau) = (\tau, \mathbf{X}(\sigma,\tau))$.

Varying the action with respect to the worldsheet coordinates and the antisymmetric tensor field yields the string equations of motion and field equations. In the temporal transverse string gauge, $X^0 = t = \tau$ and $\dot{\mathbf{X}} \cdot \mathbf{X}' = 0$, the equations of motion for the worldsheet coordinates are

$$\mu_0 \left[ \ddot{\mathbf{X}} - \frac{1}{\epsilon}\left(\frac{\mathbf{X}'}{\epsilon}\right)' \right] = \mathbf{f}, \quad \mu_0 \dot{\epsilon} = f^0, \quad (2)$$

where

$$F^\mu = (f^0, \epsilon\mathbf{f} + f^0\dot{\mathbf{X}}) = 2\pi f_a H^{\mu\alpha\beta}V_{\alpha\beta}, \quad (3)$$

and

$$\epsilon^2 = \frac{\mathbf{X}'^2}{1-\dot{\mathbf{X}}^2}. \quad (4)$$

The field equation can be simplified using the analogue of the Lorentz gauge in electromagnetism, $\partial_\mu B^{\mu\nu} = 0$. Using this gauge, the field equations are

$$\Box B_{\alpha\beta} = 2\pi f_a \int d\bar{\sigma} d\bar{\tau}\delta^4(x - X(\bar{\sigma},\bar{\tau}))V_{\alpha\beta}. \quad (5)$$

This equation can be solved easily using standard retarded Green function techniques. In analogy to electromagnetism, one can perform a series

[*]We acknowledge the support of the Particle Physics and Astronomy Research Council, in particular the Cambridge Relativity rolling grant (GR/H71550) and Computational Science Initiative grants (GR/H67652 & GR/H57585)

of manipulations to deduce the Lienard-Wiechart potential[9],

$$B_{\alpha\beta} = \frac{f_a}{2} \int d\bar{\sigma} \left( \frac{V_{\alpha\beta}}{|\Delta \cdot \dot{X}|} \right) \bigg|_{\bar{\tau}=\tau_R}, \quad (6)$$

and its derivative,

$$\partial_\mu B_{\alpha\beta} = \frac{f_a}{2} \int d\bar{\sigma} \frac{1}{\Delta \cdot \dot{X}} \frac{\partial}{\partial \bar{\tau}} \left( \frac{\Delta_\mu V_{\alpha\beta}}{|\Delta \cdot \dot{X}|} \right) \bigg|_{\bar{\tau}=\tau_R}, \quad (7)$$

where $\Delta_\mu = x_\mu - X_\mu(\bar{\sigma}, \bar{\tau})$, $\Delta^2_{\bar{\tau}=\tau_R} = 0$ and $\tau_R < t$. Using the sum and difference of the retarded and advanced time Green functions, one can calculate the self- and radiation-fields respectively. However, since the Lienard-Wiechart potential and its derivative both involve an integration along the string, which in the case of long strings is infinite, we must make some approximation in order to calculate the field at some point on the string. We shall employ the 'local backreaction approximation'[3]. In this approximation the integrals in (6) and (7) can be truncated at some scale $\Delta$, which at present is arbitrary. One can then expand the expressions as a power series in $\Delta/R$ (assuming $\Delta < R$), where $R$ is the average radius of curvature of the string.

Retaining leading order terms, one can deduce expressions for the forcing terms in the equation of motion (2) as

$$f^{0,\text{self}} = -2\pi f_a^2 \log(\Delta/\delta)\dot{\epsilon}, \quad (8)$$

$$\mathbf{f}^{\text{self}} = -2\pi f_a^2 \log(\Delta/\delta) \left[ \ddot{\mathbf{X}} - \frac{1}{\epsilon}\left(\frac{\mathbf{X}'}{\epsilon}\right)' \right], \quad (9)$$

$$f^{0,\text{rad}} = \frac{4\pi f_a^2 \Delta}{3} \left[ \frac{\epsilon^2 \dot{\mathbf{X}} \cdot \dddot{\mathbf{X}}}{1 - \dot{\mathbf{X}}^2} \right], \quad (10)$$

$$\mathbf{f}^{\text{rad}} = \frac{4\pi f_a^2 \Delta}{3} \left[ \dddot{\mathbf{X}} - \frac{1}{\epsilon}\left( \frac{\mathbf{X}' \cdot \dddot{\mathbf{X}}}{1 - \dot{\mathbf{X}}^2} \right) \mathbf{X}' \right], \quad (11)$$

where $\delta \sim f_a^{-1}$ is the string width.

The self-force can be clearly seen to be a multiple of the equations of motion[8,9], which facilitates the well-known renormalisation,

$$\mu(\Delta)\left[ \ddot{\mathbf{X}} - \frac{1}{\epsilon}\left(\frac{\mathbf{X}'}{\epsilon}\right)' \right] = \mathbf{f}^{\text{rad}}, \quad \mu(\Delta)\dot{\epsilon} = f^{0,\text{rad}}, \quad (12)$$

where $\mu(\Delta) = \mu_0 + 2\pi f_a^2 \log(\Delta/\delta)$ is the renormalised string tension. The finite radiation backreaction force experienced by the string due to $\mathbf{f}^{\text{rad}}$ and $f^{0,\text{rad}}$ is the analogue of the Abraham-Lorentz force for the point electron in classical electrodynamics.

Performing numerical simulations of the point electron using the Abraham-Lorentz force is hazardous, since the triple derivative with respect to time leads to unphysical exponentially growing solutions. However, in the case of strings one can make the approximation $\dddot{\mathbf{X}} = \ddot{\mathbf{X}}''/\epsilon^2$, which reduces the backreaction force to the analogue of a viscosity term[3]. Once one has made this approximation, methods similar to those used in ref.[6] can be used to evolve general string trajectories.

For general long string perturbations parameterized by wavelength $L$ and relative amplitude $\varepsilon = 2\pi A/L$, where $A$ is the actual amplitude, the radiation power per unit length due to the backreaction force can be seen to be

$$\frac{dP}{dl} = \frac{\beta \epsilon^2}{L}, \quad (13)$$

for some constant $\beta \sim f_a^2$. This leads to exponential decay of $\varepsilon$, since the energy per unit length in the perturbation is proportional to $\varepsilon^2$. This was verified by numerical simulations of string trajectories. We also compared the decay of string trajectories to those seen in numerical field theory. The rate of decay of amplitude was found to agree, if the correlations in the long range fields were suppressed at the curvature radius and $\Delta \sim R/4$. We also observed that an initially sharp kink visibly rounds under the action of this force.

## 3. NETWORK EVOLUTION

Our current understanding of the evolution of a cosmic string network is based on a marriage between analytic models and sophisticated network simulations[5]. However, the network simulations just evolve the free equations of motion for a string in an expanding universe. In order to incorporate the radiative effects discussed in the preceding section one must modify the equations of motion to include a radiation damping term,

$$\mu_0 \left[ \ddot{\mathbf{X}} + \frac{2\dot{a}}{a}(1 - \dot{\mathbf{X}}^2)\dot{\mathbf{X}} - \frac{1}{\epsilon}\left(\frac{\mathbf{X}'}{\epsilon}\right)' \right] = \mathbf{f}, \quad (14)$$

$$\mu_0 \left[ \dot{\epsilon} + \frac{2\dot{a}}{a}\epsilon \dot{\mathbf{X}}^2 \right] = f^0, \quad (15)$$

where $a$ is the scale factor.

However, to calculate this radiation damping term one must use Green functions in an expanding background[10]. In the radiation era, the retarded Green function is

$$D_{\text{ret}}(x, x') = \frac{a(\eta)}{2\pi a(\eta')}\delta\left((x - x')^2\right)\theta(\eta - \eta'). \quad (16)$$

where $x^\mu = (\eta, \mathbf{x})$ and $x'^\mu = (\eta', \mathbf{x}')$. Applying, the 'local backreaction approximation' in this scenario, one finds that the forcing terms are given by,

$$\mathbf{f}^{\text{rad}} = \mathbf{f}^{\text{rad}}_{\text{flat}} + \frac{\dot{a}}{a}\mathbf{g} + \mathcal{O}(1/t^3), \quad (17)$$

$$f^{0,\text{rad}} = f^{0,\text{rad}}_{\text{flat}} + \frac{\dot{a}}{a}g^0 + \mathcal{O}(1/t^3), \quad (18)$$

where the flat suffix denotes the flat space backreaction force given by (10) and (11), and $(g^0, \mathbf{g})$ is a correction to the force due to the expanding background.

The forced equations of motion (15) can be used to derive equations for the evolution of the density of long strings ($\rho_\infty$) and loops ($\rho_L$), under the influence of the expansion, Hubble damping and the radiation backreaction force. If one now inserts a term to take into account of loop production, the equations become

$$\dot{\rho}_\infty = -\frac{2\dot{a}}{a}(1 + \langle v^2 \rangle)\rho_\infty - \frac{c\rho_\infty}{L} - \frac{d\rho_\infty}{L}, \quad (19)$$

$$\dot{\rho}_L = -\frac{3\dot{a}}{a}\rho_L + \frac{c\rho_\infty}{L}, \quad (20)$$

where

$$d = d_0 + \frac{\dot{a}}{a}d_1 + \mathcal{O}(1/t^2), \quad (21)$$

$\langle v^2 \rangle$ is the average string velocity, $d_0, d_1, ..$ are constants and $c$ is a measure of the efficiency of loop production.

If one now substitutes $\rho_\infty = \mu\zeta/t^2$ and $L = \zeta^{-1/2}t$ into (19), then one can deduce a differential equation for $\zeta$. This equation has an attractive fixed point, which corresponds to the scaling regime. If $d_i = 0$ for $i > 0$, then one can deduce that

$$c = \zeta^{-1/2}(1 - \langle v^2 \rangle) - d_0. \quad (22)$$

In the case where $d_1$ is non-zero, One should observe transient effects in the scaling. However, for large times these effects will be negligible and the attractive fixed point is exactly that for $d_1 = 0$.

## 4. RADIATION CONSTRAINTS

The evolution of a network of global strings is of particular interest, since the best motivated scenario for the production of global strings is at the Peccei-Quinn phase transition. In this case, the radiated Goldstone boson is the initially massless axion, which acquires mass at the QCD phase transition, and may form a substantial part of the dark matter in the universe. In the light of the discussion of the previous section, we concluded that a network of global strings will evolve by the production of loops, which subsequently radiate into Goldstone bosons[12]. This mechanism is similar to that generally accepted to be the case for local strings. However, since the radiation into Goldstone bosons is about three orders of magnitude stronger than that into gravitational radiation, the loop production size $\alpha$ and the string scaling density $\zeta$ are likely to be different.

Using the scaling balance arguments of §3, one can deduce that the loop number density is given by[2]

$$n(l, t) \approx \frac{0.4\alpha^{1/2}\zeta(1 + \kappa/\alpha)^{5/2}}{t^{3/2}(l + \kappa t)^{5/2}}, \quad (23)$$

where $\kappa$ is the loop radiation backreaction scale. From this one can deduce that the spectral density of axions emitted by the loops is

$$\frac{\partial \rho_a}{\partial \omega}(t) \approx \frac{450 f_a^2 \alpha^{1/2}}{\omega t^2 \kappa^{3/2}}\left(1 + \frac{\kappa}{\alpha}\right)^{5/2}$$
$$\times \left[1 - \left(1 + \frac{\alpha}{\kappa}\right)^{-3/2}\right]. \quad (24)$$

Assuming axion number conservation from the QCD phase transition to the present day one can deduce that the contribution to the density of the universe from axions produced by the loops is

$$\Omega_{a,\ell} \approx 10.7 \left(\frac{\alpha}{\kappa}\right)^{3/2} \left[1 - \left(1 + \frac{\alpha}{\kappa}\right)^{-3/2}\right] h^{-2} \Delta$$
$$\times \left(1 + \frac{\kappa}{\alpha}\right)^{5/2} \left(\frac{T_0}{2.7\mathrm{K}}\right)^3 \left(\frac{f_a}{10^{12}\mathrm{GeV}}\right)^{1.18}, \quad (25)$$

where $T_0$ is the current temperature of CMBR, the Hubble constant is $H_0 = 100h\,\mathrm{km\,s^{-1}\,Mpc^{-1}}$, $\Delta \sim 10^{\pm 0.5}$ takes into account the uncertainties of the QCD phase transition and $0.1 \lesssim \alpha/\kappa \lesssim 1.0$. This contribution is substantially larger than any of the other possible contributions from long string radiation and domain wall collapse. It is also much larger than original estimates which assumed axion production via coherent oscillations of a scalar field. This difference can be understood intuitively as being due to the renormalised string energy per unit length.

## 5. DISCUSSION

We have summarized the current understanding of string network evolution and its relevance to constraints on the axion. The formalism we described in §2 offers the prospect of evolving a string network numerically, with inclusion of radiative effects. We have shown that the effect of the radiation backreaction force on the largest scales of the string network will be similar to that in the most recent analytic work[11]. However, since the Goldstone boson radiation is much stronger than gravitational radiation, realistic network simulations of global strings should see the scaling of small scale structure much earlier than for local strings.

At present the most accurate estimate of the axion density due to string loops (25) is dependent on the ratio $\alpha/\kappa$. This ratio expresses the key uncertainty in string network evolution and should in principle be calculable, to within a factor of two, in the near future.

Assuming $\Delta \approx 1$, $T_0 \approx 2.7\mathrm{K}$ and imposing the constraint $\Omega_a < 1$, one can deduce a constraint on $f_a$. If $\alpha/\kappa \approx 1$ then

$$f_a \lesssim 1.4 \times 10^{10}\mathrm{GeV}, \quad h = 0.5, \quad (26)$$

$$f_a \lesssim 4.4 \times 10^{10}\mathrm{GeV}, \quad h = 1.0, \quad (27)$$

whereas if $\alpha/\kappa \approx 0.1$ then

$$f_a \lesssim 2.5 \times 10^{10}\mathrm{GeV}, \quad h = 0.5, \quad (28)$$

$$f_a \lesssim 8.0 \times 10^{10}\mathrm{GeV}, \quad h = 1.0. \quad (29)$$

At their most extreme, the parameter uncertainties $\Delta h^{-2}[T_0/(2.7\mathrm{K})]^3$ can vary in the range 0.2 to 25, therefore for $0.1 \lesssim \alpha/\kappa \lesssim 1.0$ the constraint must lie in the region

$$f_a \lesssim 10^9 - 10^{11}\mathrm{GeV}. \quad (30)$$

Note that even the weakest string bound is stronger than the early estimate due to coherent zero-momentum axions, $f_a \lesssim 10^{12}\mathrm{GeV}$.